\documentclass[twocolumn,apl,amsmath,amssymb,showpacs,superscriptaddress]{revtex4-1}
\usepackage{epsf}      
\usepackage{graphicx}
\usepackage{color}
\usepackage{soul}
\usepackage{gensymb}
\usepackage{sidecap}
\usepackage{amsmath}
\usepackage{mathtools}

\begin{document}
\title{Observation of anisotropic thermal expansion and Jahn-Teller effect in double perovskites Sr$_{2-x}$La$_x$CoNbO$_6$ using neutron diffraction}

\author{Ajay Kumar}
\affiliation{Department of Physics, Indian Institute of Technology Delhi, Hauz Khas, New Delhi-110016, India}
\author{Anil Jain}
\affiliation{Solid State Physics Division, Bhabha Atomic Research Centre, Mumbai 400085, India}
\affiliation{Homi Bhabha National Institute, Anushakti Nagar, Mumbai 400094, India}
\author{S. M. Yusuf}
\affiliation{Solid State Physics Division, Bhabha Atomic Research Centre, Mumbai 400085, India}
\affiliation{Homi Bhabha National Institute, Anushakti Nagar, Mumbai 400094, India}
\author{R. S. Dhaka}
\email{rsdhaka@physics.iitd.ac.in}
\affiliation{Department of Physics, Indian Institute of Technology Delhi, Hauz Khas, New Delhi-110016, India}

\date{\today}      

\begin{abstract}

We use temperature dependent neutron powder diffraction (NPD) to investigate the structural changes and magnetic interactions in double perovskite Sr$_{2-x}$La$_x$CoNbO$_6$ ($x=$ 0.4 and 0.6). A structural phase transition from tetragonal ($I4/m$) to monoclinic ($P2_1/n$) is observed between $x=$ 0.4 and 0.6 samples. Interestingly, temperature evolution of the unit cell parameters follow the Gr\"uneisen approximation, and the analysis suggest the isotropic thermal expansion in case of the $x=$ 0.4 sample, whereas the $x=$ 0.6 sample shows the anisotropy where the thermal expansion along the $c$-axis significantly deviates from the Gr\"uneisen function. We observe the $z$-out Jahn-Teller distortion in the CoO$_6$ and consequently $z$-in distortion in the adjacent NbO$_6$ octahedra. With increase in the La substitution, a decrease in the degree of octahedral distortion is evident from the significant reduction in the local distortion parameter $\Delta$ around the B-site atoms. 
\end{abstract}

\maketitle
\section{\noindent ~Introduction}

Perovskite oxides (ABO$_3$, where A: rare earth/alkali earth metals, B: transition metals) have attracted great attention due to their intriguing physical properties such as half metallicity, giant magnetoresistance, multiferroicity etc. \cite{Perez_PRL_98, Kundu_PRB_05, Giovannetti_PRL_09}, resulting in their  technological applications in several fields like resistive switching devices, magnetocaloric effect, solid oxide fuel cells, photovolatics etc. \cite{Sheng_PRB_09, Das_PRB_17, Tao_NM_03, Chakrabartty_NP_18}. In this family, 50\% substitution at the B-site atoms give rise to an extra degree of freedom to tune the rock salt like alternating ordering of the BO$_6$ and B$^\prime$O$_6$ octahedra at the corners, resulting in the doubling of the unit cells in all the three crystallographic directions. A large difference in the oxidation state and large ionic mismatch between two B-site cations favor the ordered structure \cite{Anderson_SSC_93, King_JMC_10,Vasala_SSC_15, Galasso_JPC_62}, which strongly influences their magnetic, electronic and transport properties \cite{Niebieskikwiat_PRB_04, Sanchez_PRB_02, Jung_PRB_07, Erten_PRL_11, Frontera_PRB_04}. However, exact quantification of these disorders, governed by the ordered parameter, $S= $2$m$-1, where $m$ is the fractional occupancy of B/B$^\prime$ cations at their respective Wyckoff positions, is difficult from the conventional x-ray diffraction (XRD) technique, when two B-site atoms have almost similar number of electrons \cite{Saines_JSSC_07}. Further, the octahedral tilting and superstructure (doubling of the unit cell) lead to the same Bragg reflections in many cases \cite{Vasala_SSC_15}, which further limit the exact quantification of these disorders. Thus, combined neutron power diffraction (NPD) and XRD measurements can be useful for the purpose \cite{Vasala_SSC_15, Saines_JSSC_07}. Further, the degree of disorders decides several competing exchange interactions, resulting in the complex magnetic properties in these compounds, which can be effectively probed by the NPD measurements. In the perovskite family, the Co-based oxides are particularly interesting due to the various possible oxidation and spin states of Co in the octahedral coordination, which is governed by the charge neutrality, and the competing Hund's exchange energy and crystal field splitting, respectively. Larger crystal field in case of Co$^{3+}$ increase the possibility of the spin-paring and hence stabilize the low spin (LS) state (3$d^6$; t$_{2g}^6$e$_g^0$) as compared to the Co$^{2+}$ \cite{Kumar_PRB1_20, Bos_PRB_04}. On the other hand, due to the weak crystal field, Co$^{2+}$ is widely know to preserve the free Co$^{2+}$ like $^4T_1$ ground term, resulting in the large unquenched orbital moment and lead to the long-range antiferromagnetic (AFM) interactions \cite{Lloret_ICA_08}. Further, several competing interactions due to various possible valence and spin states of cobalt result in the frustration between the spins, which give rise to the low temperature glassy behavior in the complex oxides \cite{Shukla_PRB_18, Shukla_JPCC_19, Kumar_PRB2_20}, where investigation using NPD plays an important role \cite{Sow_PRB_12}. Thus, it is interesting to investigate the magnetic and structural properties of the Co based perovskites with change in the B-site ordering, valance states of Co and hence the resulting the crystal field.

In this context, we have recently studied the correlation between structural and magnetic properties of Sr$_{2-x}$La$_x$CoNbO$_6$ ($x=$ 0--1), where substitution of each La$^{3+}$ ion at Sr$^{2+}$ site convert one Co ion from 3+ to the 2+ state and therefore, a significant reduction in the crystal field is expected. We find the enhancement in the B-site ordering, AFM interactions and insulating nature of the samples with the La substitution \cite{Kumar_PRB1_20, Kumar_PRB2_20}. The magnetization measurements suggest the possible combination of different spin-states of the Co$^{3+}$ along with the HS Co$^{2+}$ in these compounds. Further, the non-linearity in the inverse susceptibility curves suggest the possibility of the temperature dependent spin-state transition in $x\leqslant0.4$ samples. On the other hand, the microscopic analysis by the specific heat measurements indicate the persistence of the discrete energy states resulting from the crystal field splitting, spin orbit coupling and octahedral distortion \cite{Kumar_PRB2_20}. However, extant of the octahedral distortion and its evolution with the temperature and La substitution have not been quantified, which can be the key factor in governing these properties. Also, the detailed identification of the origin of the structural and magnetic transition occur from tetragonal to monoclinic and cluster glass to AFM from $x=$ 0.4 to 0.6 samples is required to understand their exotic physical properties.

Therefore, in this paper we use temperature dependent NPD to probe  the crystallographic and magnetic structure of Sr$_{2-x}$La$_x$CoNbO$_6$ ($x= $0.4 and 0.6) samples across the transition. Evolution of the superlattice reflection clearly evident the B-site ordered structure in case of the $x=$0.6 sample. The Rietveld refinement  of the NPD data indicate the structural transition from tetragonal ($I4/m$) to monoclinic ($P2_1/n$) phase from $x=$ 0.4 to 0.6. The temperature evolution of the unit cell volume follow the  Gr\"uneisen approximation for both the samples; however, the other unit cell parameters ($a$, $b$, and $c$) of the $x=$ 0.6 sample show the anisotropic thermal expansion. The distortion in (Co/Nb)O$_6$ octahedra is found to decrease with the La substitution from $x=$ 0.4 to 0.6 due to large concentration of the Jahn-Teller (JT) inactive Co$^{2+}$ ions in the latter case. Unexpectedly, the absence of the magnetic reflections down to 2~K suggests the disorder induced short range magnetic correlations in both these $x=$ 0.4 and 0.6 samples.

\section{\noindent ~Experimental}

Polycrystalline samples of Sr$_{2-x}$La$_{x}$CoNbO$_{6}$ ($x=$ 0.4 and 0.6) were synthesized by usual solid-state route. Details of the synthesis process have been given elsewhere \cite{Kumar_PRB1_20}. The temperature dependent NPD data have been measured using the powder diffractometer PD-I ($\lambda= $1.094\AA) and PD-II ($\lambda= $1.2443\AA) at the Dhruva reactor, Trombay, Mumbai, India. For the Rietveld refinement of NPD patterns, we use the pseudo-voigt peak shape and linear interpolation between the set background points in FullProf suite \cite{Carvajal_PB_93}. The room temperature NPD pattrens for both the samples were fitted with the nuclear phase only and then the respective low temperature pattrens were fitted by fixing all the fitting parameters and allowing only oxygen position and lattice parameters to vary with the temperature. The neutron depolarization measurements were performed on $x=$ 0.4 sample from 1.5 to 300~K using the polarized neutron spectrometer (PNS) ($\lambda=$ 1.205\AA) at the Dhruva reactor, Trombay, Mumbai, India. The measurements were performed in the warming mode at 50~Oe applied magnetic field, after cooling the sample to 1.5~K in the same field. The incident and transmitted neutron beams were polarized and measured using Cu$_2$MnAl(100) (along -$z$ direction) and Co$_{0.92}$Fe$_{0.2}$(200) (along +$z$ direction) crystals, respectively. The intensity ratio of the neutron beam, when $\pi$ flipper (placed before the sample) is off and on, defined as the flipping ratio(R), was measured as a function of sample temperature \cite{Yusuf_Pramana_96}.

\section{\noindent ~Results and discussion}

\begin{figure}
\includegraphics[width=3.7in]{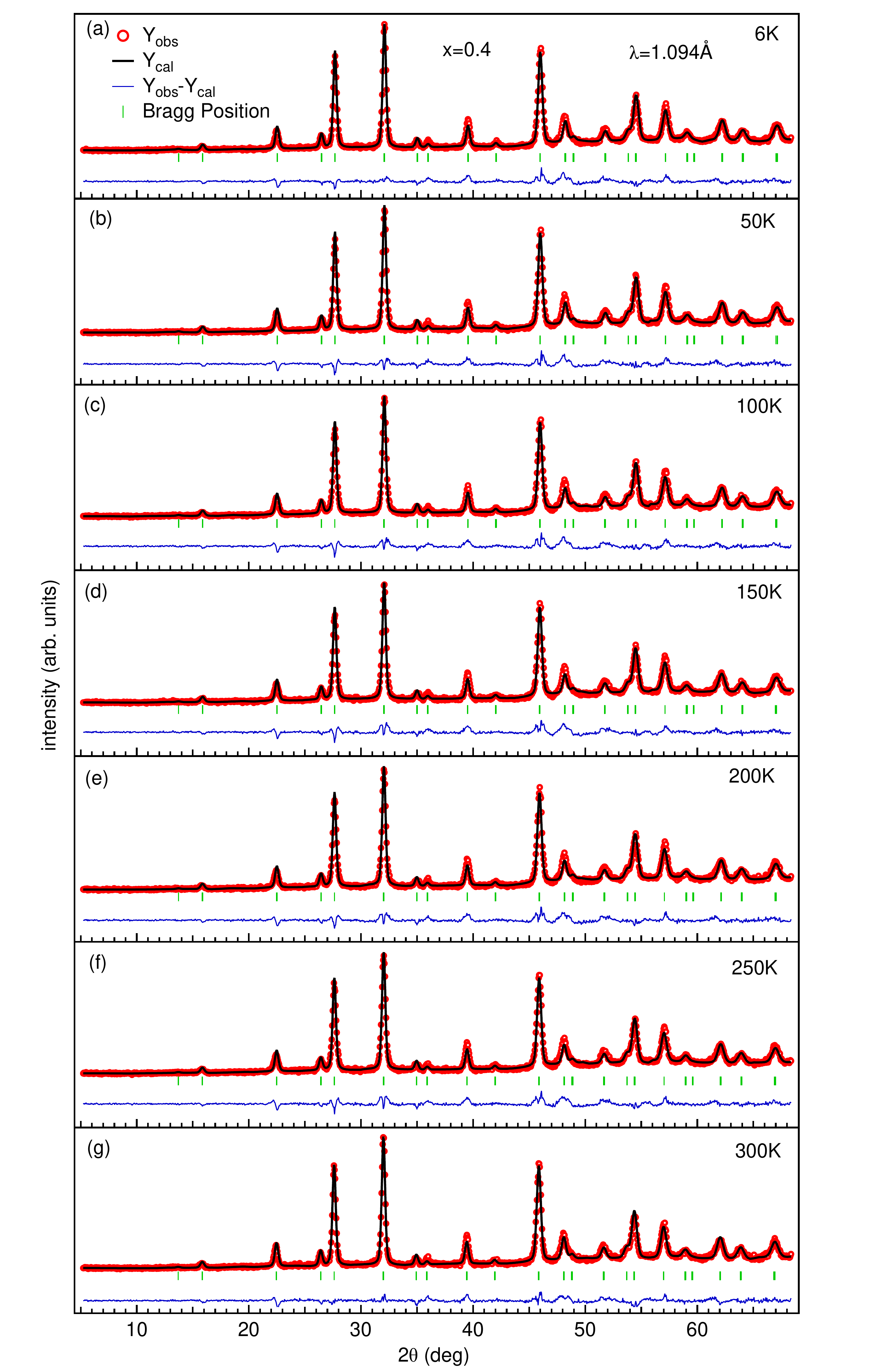}
\caption {. (a)--(g) The Rietveld refinement of NPD data of $x= $0.4 sample recorded with $\lambda=$  1.094\AA~ from 6~K to 300~K, where the observed data points, simulated curve, Bragg peak positions, and difference in observed and simulated data are shown by open red circles, a continuous black line, vertical green bars, and a continuous blue line, respectively.}
 \label{all_4}
\end{figure}

\begin{figure} 
\includegraphics[width=3.75in]{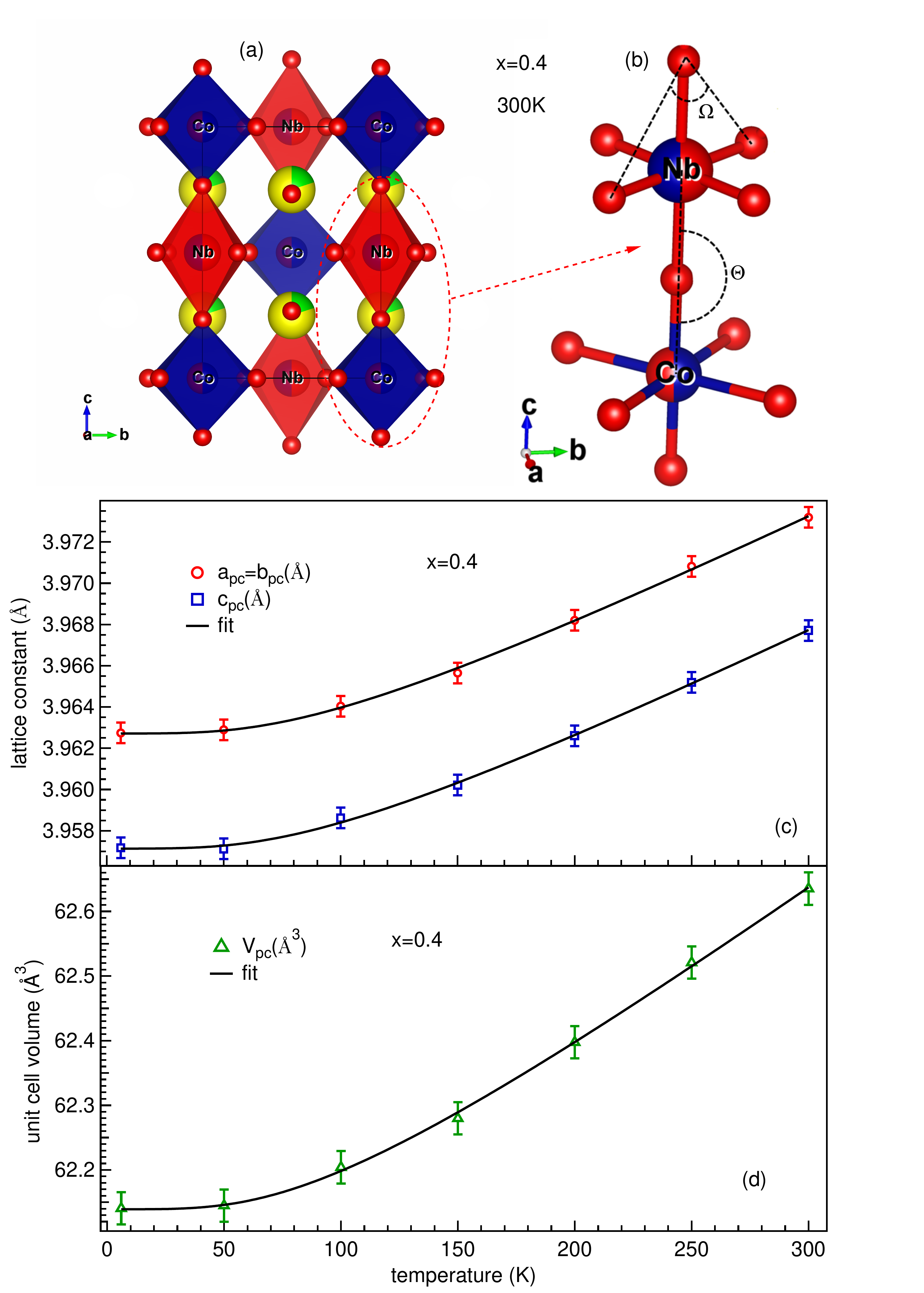}
\caption {(a) Tetragonal crystal structure of the $x=$ 0.4 sample at 300~K as viewed from the $a$-axis. (b) Two representative octahedra to show the octahedral distortion parameters, $\Theta$ and $\Omega$. (c) The temperature evolution of the pseudocubic lattice parameters ($a_{\rm pc}$=$a$/$\sqrt{2}$=$b_{\rm pc}$=$b$/$\sqrt{2}$, $c_{\rm pc}$=$c$/2), and (d) the unit cell volume, extracted from the Rietveld refinement of the NPD patterns of the $x=$ 0.4 sample from 6--300~K. The solid black lines represent the best fit using the Gr\"uneisen approximation.} 
\label{fit_4}
\end{figure}

\begin{table}
	
		\label{Table_fit}
		\caption{Rietveld refinement parameters for the $x=$ 0.4 and 0.6 samples at 300~K. We use $I4/m$ space group, Nb/Co at 2$b$(0, 0, $\frac{1}{2}$), Co/Nb at 2$a$(0, 0, 0), O1 at 4e(0, 0, z) with 0.25 occupancy and O2 at 8$h$ (x, y, 0) with 0.5 occupancy for the $x=$ 0.4 sample, and $P2_1/n$ space group, La/Sr, O1, O2, O3 at 4$e$($x$, $y$, $z$), Nb/Co at 2$d$($\frac{1}{2}$, 0, 0) and Co/Nb at 2$c$($\frac{1}{2}$, 0, $\frac{1}{2}$) for the $x=$ 0.6 sample.}
		
		\begin{tabular}{p{2.7cm}p{2.7cm}p{2.7cm}}
\hline
		\hline
	 & 0.4 & 0.6\\
\hline
           La/Sr x & 0 & 0.0023(4) \\
	     y & 0.5 & 0.0123(8) \\
	     z &0.25 &0.2522(6) \\
          100 x U (\AA$^2$) & 0.91(2) & 1.04(3)\\

          Nb/Co occ & 0.502(3)/0.498(3) & 0.842(6)/0.158(6) \\
	     100 x U  & 0.98(1) & 1.12(6) \\

	     Co/Nb occ & 0.502(3)/0.498(3) & 0.842(6)/0.158(6) \\
	     100 x U  & 0.98(1) & 1.12(6)\\

          O1 x & 0 & 0.2680(1) \\
	     y & 0 & 0.2754(7) \\
	     z &0.2330(3) &0.0385(4)\\
          100 x U (\AA$^2$) & 0.72(3) & 0.69(4) \\

          O2 x & 0.2409(2) & 0.2362(5) \\
	     y & 0.3039(4) & -0.2331(1) \\
	     z & 0 &0.0151(7)\\
          100 x U (\AA$^2$) & 0.81(2) &0.93(7)\\

         O3 x & - & -0.0621(8) \\
	     y & - & 0.4994(7) \\
	     z & - &0.2599(6)\\
          100 x U (\AA$^2$) & - & 1.02(4)\\

\hline
\hline
\end{tabular}
\end{table}

We present the Rietveld refinement of the NPD patterns for the $x=$ 0.4 sample from 6--300~K in Figs.~\ref{all_4}(a--g). There are no additional Bragg reflections down to 6~K, which indicate the absence of any long-range magnetic ordering in this temperature range. The NPD patterns for all the temperatures were best fitted using the nuclear phase only with the tetragonal ($I4/m$) structure having a$^0$a$^0$c$^-$ tilt and random occupancy of the Co and Nb atoms at the B-sites ($S=$ 0), as shown in Fig.~\ref{fit_4}(a) and the refined atomic parameters at 300~K are presented in Table I. Further, we extract the temperature evolution of the pseudocubic lattice parameters $a_{\rm pc}$=$a$/$\sqrt{2}$=$b_{\rm pc}$=$b$/$\sqrt{2}$, $c_{\rm pc}$=$c$/2, and the corresponding unit cell volume for the $x=$ 0.4 sample from the refinement and plotted in Figs.~\ref{fit_4}(c, d), respectively. The thermal expansion of the unit cell is evident from the Figs.~\ref{fit_4}(c, d) due to an increase in the lattice vibrations. We found that the unit cell expands at the relatively slower rate at the lower temperatures as compared to that at the higher temperatures [see Figs.~\ref{fit_4}(c, d)]. Here, it is important to note that due to the insulating nature of these samples \cite{Kumar_PRB1_20}, the electronic part in the thermal expansion is expected to be negligible as compared to the phononic contribution. Therefore, the thermal evolution due to the phonons can be approximated by the Gr\"uneisen function up to the first order as \cite{Wallace_book_98, Zhu_PRB_20}
\begin{equation} 
\epsilon(T)=\epsilon_0 + K_0U, 
\label{approx}
\end{equation} 
where $\epsilon_0$ is the lattice configuration at 0~K, $K_0$ is the measure of the incompressibility of the sample, and $U$ is the internal energy which can be approximated using the Debye model as,
\begin{equation}
U(T) = 9Nk_BT\left(\frac{T}{\theta_D}\right)^3 \int_0^{\theta_D/T} \frac{x^3}{(e^x-1)}dx, 
\label{Debye}
\end{equation} 
where $N$ is the number of atoms per formula unit (ten in the present case), $k_B = $1.38062~x~10$^{-23}$ J/K is the Boltzmann constant, and $\theta_D$ is the Debye temperature. The best fit parameters using the above equations are listed in Table II and the resulting curves are shown by the black solid lines in Figs.~\ref{fit_4}(c, d) for the $x=$ 0.4 sample. Here, the temperature evolution of all the three unit cell parameters can be well approximated by the Gr\"uneisen function. However, the estimated value of the Debye temperature [353(5)~K] is much lower than that extracted from the specific heat measurement [678(8)~K], considering the contribution due to both acoustic (69\%) and optical (31\%) phonons using combined Debye and Einstein models \cite{Kumar_PRB2_20}, which indicate the non-negligible contribution from the optical phonons in these samples. Further, the incompressibility constant $K_0$ is comparable for $a=b$ and $c$ axis, indicating the isotropic thermal expansion for the $x=$ 0.4 sample.

\begin{table}
	
		\label{Table_fit}
		\caption{The parameters extracted from the Gr\"uneisen approximation of the unit cell parameters in the pseudocubic representation for the $x=$ 0.4 and 0.6 samples.}

		\begin{tabular}{p{2cm}p{2cm}p{2cm}}
\hline
		\hline
	 & 0.4 & 0.6\\
\hline
          $\epsilon_0^a$ (\AA) & 3.963(2) & 3.970 (2) \\
	  $\epsilon_0^b$ (\AA) & 3.963(2) & 3.955(3) \\
	  $\epsilon_0^c$ (\AA) &3.957(1)&-\\
          $\epsilon_0^V$ (\AA$^3$) & 62.14(2) &62.54(2)\\
          $K_0^a$ (\AA/J) &1.35 x 10$^{17}$&1.26 x 10$^{17}$\\	
          $K_0^b$ (\AA/J) &1.35 x 10$^{17}$&2.97 x 10$^{17}$\\
          $K_0^c$ (\AA/J) &1.36 x 10$^{17}$&-\\	
          $K_0^V$ (\AA$^3$/J) & 6.40 x 10$^{18}$&4.79 x 10$^{18}$\\
          $\theta_D^a$ (K)&353(5)&413(8)\\	
          $\theta_D^b$ (K)&353(5)&808(14)\\
          $\theta_D^c$ (K)&353(5)&-\\	
          $\theta_D^V$ (K)&353(5)&314(5)\\

\hline
\hline
\end{tabular}
\end{table}

\begin{figure}[ht]
\includegraphics[width=3.7in]{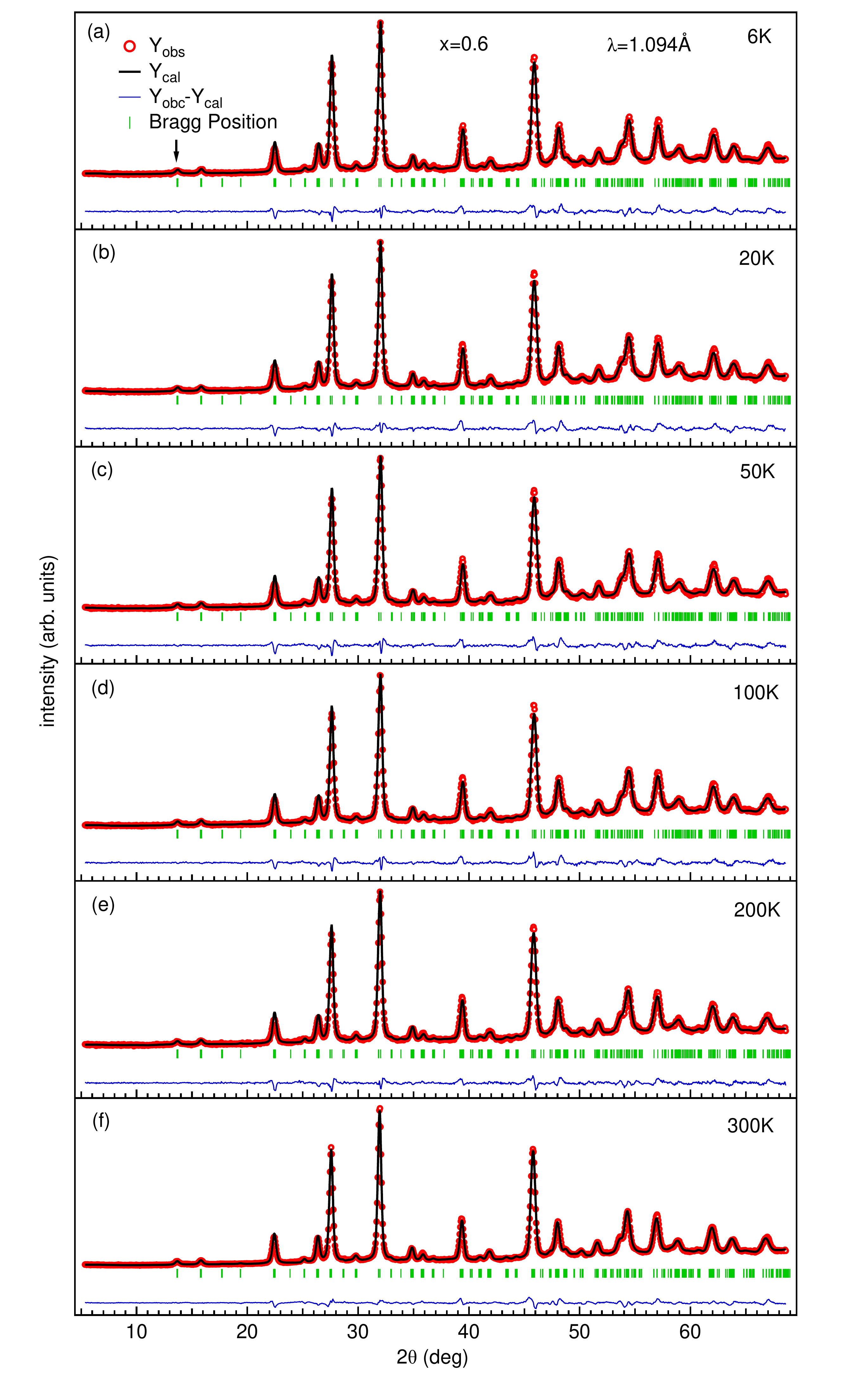}
\caption {(a--f) The Rietveld refinement of NPD patterns of the $x=$ 0.6 sample measured at $\lambda = $1.094~\AA~ from 6 to 300~K. A vertical arrow in panel (a) represent the presence of (101) superlattice reflection.} 
\label{all_6}
\end{figure}

\begin{figure} [h]
\includegraphics[width=3.5in]{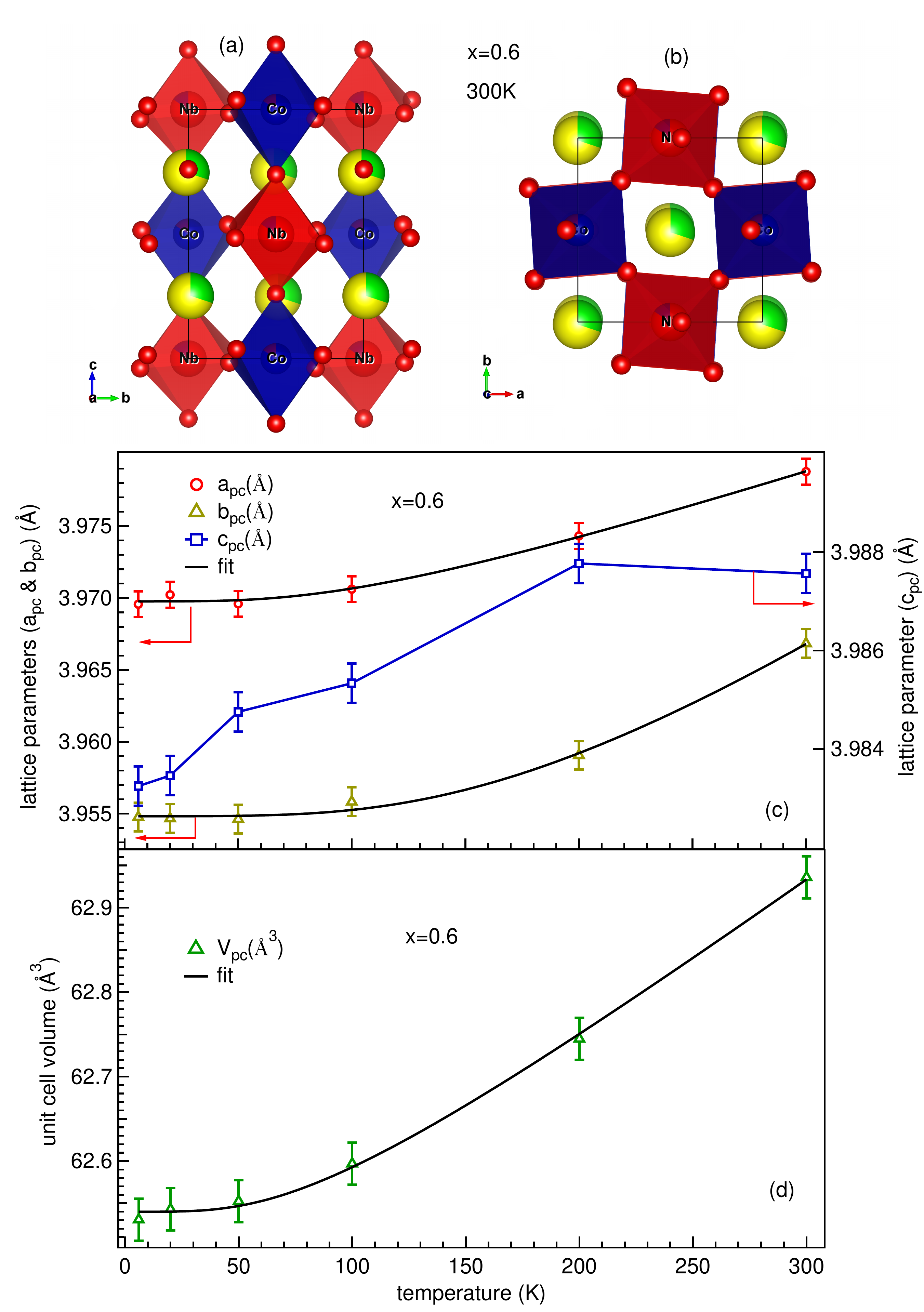}
\caption {Monoclinic crystal structure of $x =$0.6 sample at 300~K as viewed from (a) $a$-axis and (b) $c$-axis. (c) The temperature evolution of the pseudocubic lattice parameters [$a_{\rm pc}$=$a$/$\sqrt{2}$, $b_{\rm pc}$=$b$/$\sqrt{2}$ (on left axis) and $c_{\rm pc}$=$c$/2 (on right axis)], and (d) the unit cell volume, extracted from the Rietveld refinemnt of the NPD patterns of the $x=$ 0.6 sample from 6--300~K. The solid black lines represent the best fit using the Gr\"uneisen approximation.} 
\label{fit_6}
\end{figure}

In order to further understand the structural and magnetic transition in Sr$_{2-x}$La$_x$CoNbO$_6$ samples, temperature dependent NPD patterns have been recorded for the $x=$ 0.6 sample, as shown in Figs.~\ref{all_6}(a--f). Evolution of the (101) superlattice reflection can be clearly seen for all the temperatures, resulting from the enhancement of the alternating Co/Nb ordering, as shown by the vertical arrow in Fig.~\ref{all_6}(a). A close inspection of the NPD patterns during the Rietveld refinement indicate the small degree of the monoclinic distortion in the sample and the NPD patterns were fitted with the $P2_1/n$ symmetry having b$^-$b$^-$c$^+$ tilt, as shown in Figs.~\ref{fit_6}(a, b) as viewed from the $a$ and $c$-axis, respectively, and the resulting unit cell parameters are presented in Figs.~\ref{fit_6}(c, d) in the pseudocubic representation, i.e., $a_{\rm pc}$=$a$/$\sqrt{2}$, $b_{\rm pc}$=$b$/$\sqrt{2}$, $c_{\rm pc}$=$c$/2, for all the temperatures. The atomic refinement parameters at 300~K are presented in Table I. The Rietveld refinement of the NPD patterns indicate 68\% ordering between the Co/Nb atoms and pure monoclinic phase in the $x=$ 0.6 sample. It is interesting to note here that the temperature evolution of the unit cell parameter $c_{\rm pc}$ first increases rapidly and then with a much slower rate for $T>$ 100~K, which do not follow the Gr\"uneisen approximation functions, whereas the cell parameters $a_{\rm pc}$ and $b_{\rm pc}$ follow the Gr\"uneisen function, indicating the anisotropic thermal expansion of the unit cell [see Fig.~\ref{fit_6}(c)]. Further, the nature of the thermal expansion of $a_{\rm pc}$ and $b_{\rm pc}$ parameters is different, which can be observed from change in the onset of the upturn temperature in the two cases [see Fig.~\ref{fit_6}(c)]. This gives significantly different Debye temperatures in the two cases extracted from the fitting with the approximated Gr\"uneisen function, as presented in Table II. A significantly high value of the Debye temperature $\theta_D$ and incompressibility constant $K_0$ in the $b$ direction further indicate the anisotropic stiffness of the sample in the different crystallographic directions. The Debye temperature extracted from the thermal expansion of the unit cell volume, which is the cumulative effect of the all axis, is close to that for $x=$ 0.4 sample with the isotropic thermal expansion. Also, the increase in the unit cell volume with the La substitution is evident even after the substitution of the smaller La$^{3+}$ cations (1.36 \AA; 12 coordinated) at larger Sr$^{2+}$ (1.44 \AA; 12 coordinated) site [see Figs.~\ref{fit_4}(d) and \ref{fit_6}(d)], which is consistent with the reported XRD results and can be understood on the basis of the enhancement in the  concentration of bigger Co$^{2+}$ (0.650 \AA $\space$ for LS and 0.745 \AA $\space$ for HS; 6 coordinated) as compared to Co$^{3+}$ (0.545 \AA $\space$ for LS and 0.610 \AA $\space$ for HS; 6 coordinated) \cite{Kumar_PRB1_20, Shannon_AC_76}.

\begin{figure}
\includegraphics[width=3.6in]{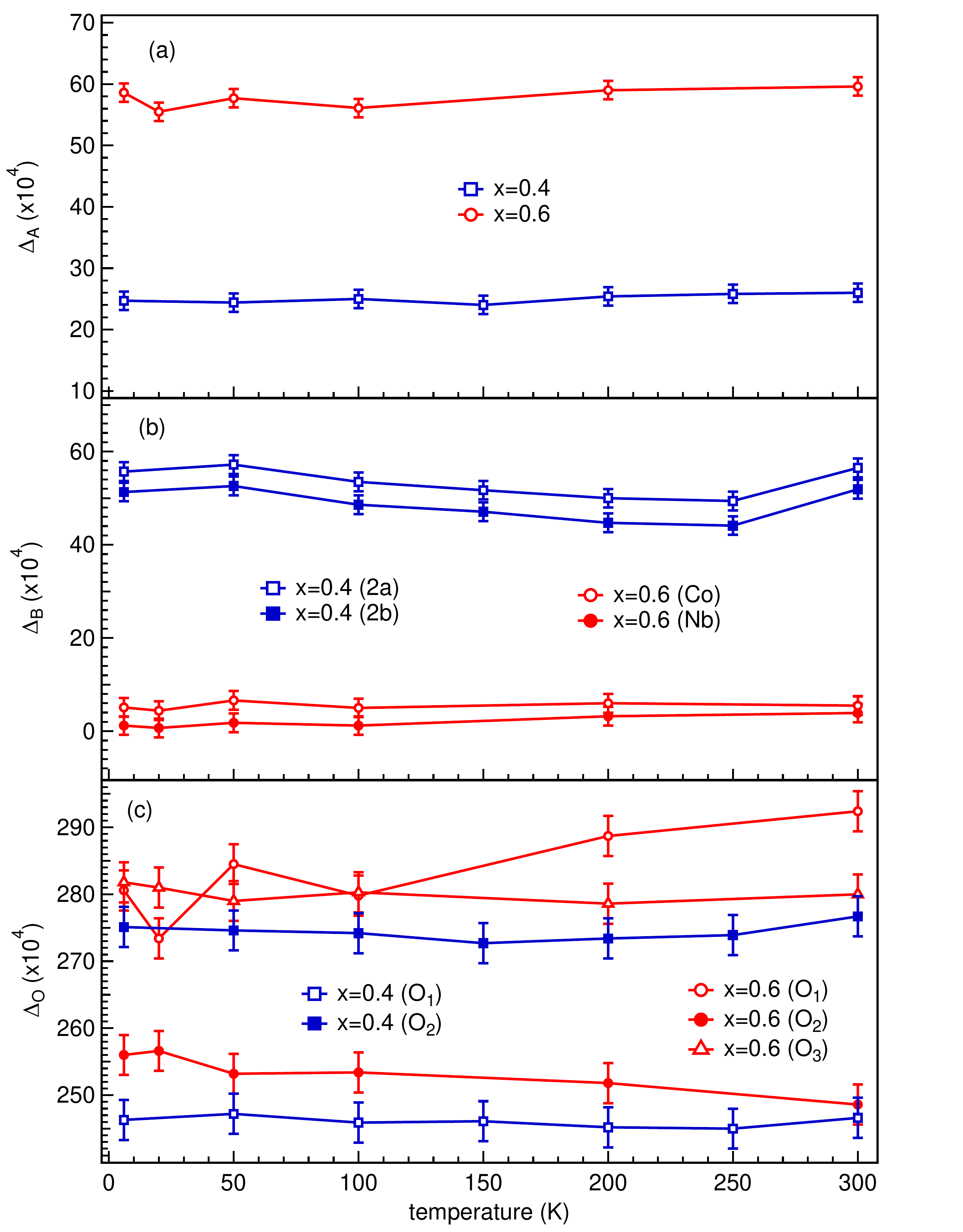}
\caption {The local distortion parameter ($\Delta$), defined by equation~\ref{distortion} around (a) $A$-site (La/Sr), (b) $B$-site (Co/Nb), and (c) oxygen atoms for the $x=$ 0.4 and 0.6 samples, as a function of temperature.} 
\label{Fig3_new}
\end{figure}

Now we discuss the behavior of extracted parameters from the above analysis for both the smaples; for example, the bond distances between cobalt and oxygen atoms are particularly important to understand the extent of Co 3$d$-O 2$p$ orbital hybridization and the resulting crystal field strength in these compounds, which is the key factors in determining the various spin-states of cobalt and consequently their complex magnetic properties. For example, Chen $\it {et} $\space$ \it{al.}$ showed that $\langle$Co--O$\rangle >$  1.93 \AA $\space$ favors the HS state of Co$^{3+}$ in the octahedral coordination, whereas an average Co--O bond length shorter than that stablize the LS state due to the large crystal field splitting as compared to the Hund's exchange energy in the latter \cite{Chen_JACS_14}. Here, the spin-orbit coupling split the t$_{2g}$ state of Co$^{3+}$ with S=2 and \~ L=1 into ground state triplet (\~ J=1) and first and second excited states with \~ J=2 and \~ J=3, respectively. Our recent study of the Schottky anomaly present in the specific heat curves of $x\leqslant$0.4 samples evident the further splitting of this spin-orbit triplet with \~ J=1 into a low lying singlet and an excited doublet due to the octahedral elongation (O$_{4h}$) \cite{Kumar_PRB2_20}. This $z$-out Jahn-Teller (JT) distortion in the CoO$_6$ octahedra and consequently $z$-in distortion in NbO$_6$ octahedra is also evident from the presence of the two sets of B--O bond distances in the $x =$ 0.4 sample, with two longer and four shorter bonds in one octahedra (at $2a$ site) and two shorter and four longer in other (at $2b$ site). The presence of the JT distortion has been also confirmed in several other perovskite cobaltites including the isostructural Sr$_2$FeCoO$_6$ double perovskite, which gives the valuable information about the energy level splitting in these compounds due to the structural instability and hence their complex magnetic, electronic and transport properties \cite{Shukla_JPCC_21, Pradheesh_EPJB_12, BaiIC19}. However, due to the random occupancy of the Co and Nb atoms in the $x=$ 0.4 sample, it is difficult to assign their position in the lattice and hence nature of the JT distortion in each octahedra. In the $I4/m$ symmetry, oxygen atoms are located at the two different lattice sites 4$e$ (0, 0, $z$) and 8$h$ ($x$, $y$, 0), named as O1 and O2, respectively, where  each B-site atom (Co/Nb) is connect with the two O1 and four O2 oxygen atoms, forming the corner shared octahedra. However, each A-site atom (Sr/La) is connected to the four O1 and eight O2 atoms, where La/Sr--O2 form two set of the equal bond lengths. The average Sr/La--O distance increases with the temperature as presented in Table III due to increase in the thermal assisted displacement of the atoms about their mean position. Further, the local distortion around each atom is important to understand the lattice deformation with the temperature as well as chemical pressure due to the La substitution. The local distortion around an atom can be quantified as \cite{Shukla_JPCC_19, Zhu_PRB_20} 
\begin{equation}
\Delta = \frac{1}{n}\sum_{i=1}^n\left(\frac{d_n-\langle d\rangle}{\langle d\rangle}\right)^2
\label{distortion}
\end{equation}
where $n$ is the number of atoms coordinated to that site, $d_n$ is the bond length along any one coordination, and $\langle d\rangle$ is the average bond length of all the coordinations of that site. It is important to note that distortion parameter $\Delta$ around the A-site atoms is significantly lower than that around B-site atoms for the $x=$ 0.4 sample, as shown in Figs.~\ref{Fig3_new}(a, b) due to the presence of the JT distortion around B-sites. We observe the highest distortion around the oxygen atoms due to significantly different La/Sr--O (four) and Co/Nb--O (two) bond distances around them [see Fig.~\ref{Fig3_new}(c)]. The temperature evolution of the distortion parameter show the non-monotonic behavior around all the lattice sites.

\begin{table*}
	
		\label{Table_fit}
		\caption{Selected bond lengths, monoclinic angle ($\beta$), and fitting reliability parameters extracted from the Rietveld refinement of the NPD data for $x = $0.4 and 0.6 samples. In addition, the tetragonal strain ($e_t$), and octahedral distortion parameters $\phi_1$ and $\phi_2$ are tabulated for the $x=$ 0.4 sample (see text for more details).}

\begin{tabular}{p{4cm}p{1.9cm}p{1.9cm}p{1.9cm}p{1.9cm}p{1.9cm}p{1.9cm}p{1.9cm}p{1.9cm}p{1.9cm}p{1.9cm}}
\hline
\hline
 & & & $x = $0.4 & & &&\\
\hline
 &6K&50K&100K&150K& 200K&250K&300K\\
\hline
            $\langle$A--O$\rangle_{12}$ (\AA) &2.8091(5)&2.8092(7) & 2.8100(6) & 2.8107(6) & 2.8126(5) & 2.8142(7) &2.8169(6) \\ 
	       (Co/Nb--O1)$_{2a}$ (\AA)[2] & 1.847(9) & 1.842(9) & 1.849(11) & 1.850(12) & 1.856(11) & 1.860(11) & 1.849(8)\\
                (Co/Nb--O2)$_{2a}$ (\AA)[4] & 2.174(9) & 2.173(10) & 2.169(10) & 2.164(13) & 2.165(10) & 2.168(13) & 2.179(9)\\

                 (Co/Nb--O1)$_{2b}$ (\AA)[2] & 2.110(9) & 2.115(9) & 2.110(11) & 2.110(12) & 2.107(11) & 2.105(11) & 2.119(8)\\
                (Co/Nb--O2)$_{2b}$ (\AA)[4] & 1.819(10) & 1.820(10) & 1.826(11) & 1.830(15) & 1.834(11) & 1.834(15) & 1.825(10)\\
    $c_{\rm pc}$/$a_{\rm pc}$  & 0.9986 & 0.9985 & 0.9986 & 0.9986 & 0.9986 & 0.9986 & 0.9986\\
 $e_t$ x 10$^{-4}$  & -7.0 & -7.3 & -6.8 & -6.9 & -7.1 & -7.1 & -6.9\\

     $\phi_1$ (deg) & 4.7 & 4.7 & 4.7 & 4.6 & 4.7 & 4.8 & 4.8\\
($\phi_2$)$_{2a}$ (deg) & -4.6 & -4.7 & -4.6 & -4.4 & -4.4 & -4.4 & -4.7\\
($\phi_2$)$_{2b}$ (deg) & 4.2 &  4.3 & 4.2 & 4.1 & 4.0 & 4.0 & 4.2\\
        R$_p$ & 3.90\% & 4.12\% & 4.42\% & 4.35\% & 4.23\% & 4.23\% & 4.63\%\\
        R$_{wp}$ & 5.20\% & 5.45\% & 5.80\% & 5.76\% & 5.60\% & 5.57\% & 4.84\%\\
        R$_F$ & 5.86\% & 5.67\% & 6.30\% & 6.95\% & 6.93\% & 6.65\% & 4.41\%\\

            \hline
            & & & $x = $0.6 & & &&\\
            \hline
           &6K&20K&50K&100K& 200K&300K&\\
           \hline
             $\langle$A--O$\rangle_{12}$ (\AA) & 2.8159(2) &  2.8162(1) & 2.8162(1) & 2.8170(3) & 2.8187 (1) & 2.8219(2) \\
	 (Co--O1) (\AA) [2] & 2.006(14) &  2.018(17) & 2.004(17) & 1.998(17) & 1.982(12) & 1.992(12)) \\
      (Co--O2) (\AA) [2] & 1.993(17) &  1.986(17) & 1.979(17) & 1.995(17) & 2.000(12) & 2.007(12) \\
      (Co--O3) (\AA) [2] & 2.096(12) &  2.088(14) & 2.100(12) & 2.093(13) & 2.095(12) & 2.100(10) \\

       (Nb--O1) (\AA) [2] & 2.003(15) &  1.984(2) & 2.009(17) & 2.011(17) & 2.041(12) & 2.043(12) \\
       (Nb--O2) (\AA) [2] & 1.993(17) &  2.002(17) & 2.003(17) & 1.987(17) & 1.985(12) & 1.981(12) \\
       (Nb--O3) (\AA) [2] & 1.952(12) &  1.962(14) & 1.949(12) & 1.957(13) & 1.955(12) & 1.948(10) \\
            
            $\beta$(deg) & 89.67(1) & 89.67(1) & 89.64(1) & 89.66(1) & 89.65(1) & 89.72(1)\\
        R$_p$ & 2.59\% & 2.63\% & 2.56\% & 2.47\% & 2.24\% & 1.76 \%\\
        R$_{wp}$ & 3.47\% & 3.52\% & 3.43\% & 3.30\% & 3.01\% & 2.44 \%\\
        R$_F$ & 5.05\% & 4.93\% & 4.83\% & 4.77\% & 3.51\% & 2.97 \%\\
            \hline
\hline
\end{tabular}
\end{table*}

Another important parameter is the $c_{\rm pc}$/$a_{\rm pc}$ ratio, which is the measure of the extent of the tetragonal distortion in the compound from the cubic symmetry. A small deviation in $c_{\rm pc}$/$a_{\rm pc}$ ratio from the unity as given in Table III indicates only a slight distortion from the cubic symmetry, which is supported by the presence of weak Raman active modes in the compound in Ref.~\cite{Kumar_PRB1_20}. Further, the tetragonal strain defined as $e_t$=($c_{\rm pc}$-$a_{\rm pc}$)/($c_{\rm pc}$+$a_{\rm pc}$) is also given in the Table III, which is close to the values reported for the Sr$_2$FeCoO$_6$ \cite{Pradheesh_EPJB_12}. The octahedral distortion can be further quantified by the degree of tilting from the $z$-axis as $\phi_1$=(180-$\Theta$)/2, where $\Theta$ is the Co(Nb)--O2--Nb(Co) bond angle along $z$-axis. Similarly, the rotation angle $\phi_2$, defined as $\phi_2$=(90-$\Omega$)/2, where $\Omega$ is O2--O1--O2 angle is also the measure of the octahedral distortion, as shown in Fig.~\ref{fit_4}(b) and the values are given in Table III. 

Moreover, in the $P2_1/n$ symmetry for the $x=$ 0.6 sample, the oxygen atoms have three different lattice positions ($4e$) named as O1, O2, and O3 with twelve different A--O bond distances, average value of which is given in Table III, which increase with the temperature due to the lattice expansion. All the A--O bonds of different bond length in the $P2_1/n$ (monoclinic) symmetry in case of the $x=$ 0.6 sample result in the higher local distortion ($\Delta$) around the A-site as compared to the $x=$ 0.4 sample, as shown in Fig. \ref{Fig3_new}(a). Further, each B-site atom has two equal bond distances with each oxygen atom forming the corner shared octahedra. Here, it is clearly evident from the Table III that the CoO$_6$ octahedra show the $z$-out JT distortion and consequently NbO$_6$ shows $z$-in, which is consistent with the energy level splitting scheme, extracted from the specific heat measurements of these compounds \cite{Kumar_PRB2_20}. Further, a slight difference in the B--O bond distances in the $xy$ plane is resulting from the small monoclinic distortion. Here, it is important to note that the degree of JT distortion (difference in B-O  bond distances in $xy$ plane with that along $z$-axis) decreases with the La substitution which is possibly due to enhancement in the Co$^{2+}$ concentration, which is less favorable for the JT distortion as compared to Co$^{3+}$. This reduction in JT distortion is more clearly reflected from the decrease in the value of local distortion parameter around B-site atoms, as presented in Fig.~\ref{Fig3_new}(b). The reduction in the octahedral distortion with the La substitution is also observed in our recent local structural investigations of the Sr$_{2-x}$La$_x$CoNbO$_6$ ($x =$0--1) samples using the extended x-ray absorption fine structure (EXAFS) measurements \cite{Kumar_XAS_22}. Further, the average distortion around oxygen atoms increases with the La substitution due to reduction in the lattice symmetry [see Fig.~\ref{Fig3_new}(c)]. The local distortion around all the atoms show the non-monotonic behavior with the temperature analogous to the $x=$ 0.4 sample, as presented in Figs.~\ref{Fig3_new}(a--c).

\begin{figure} 
\includegraphics[width=3.6in]{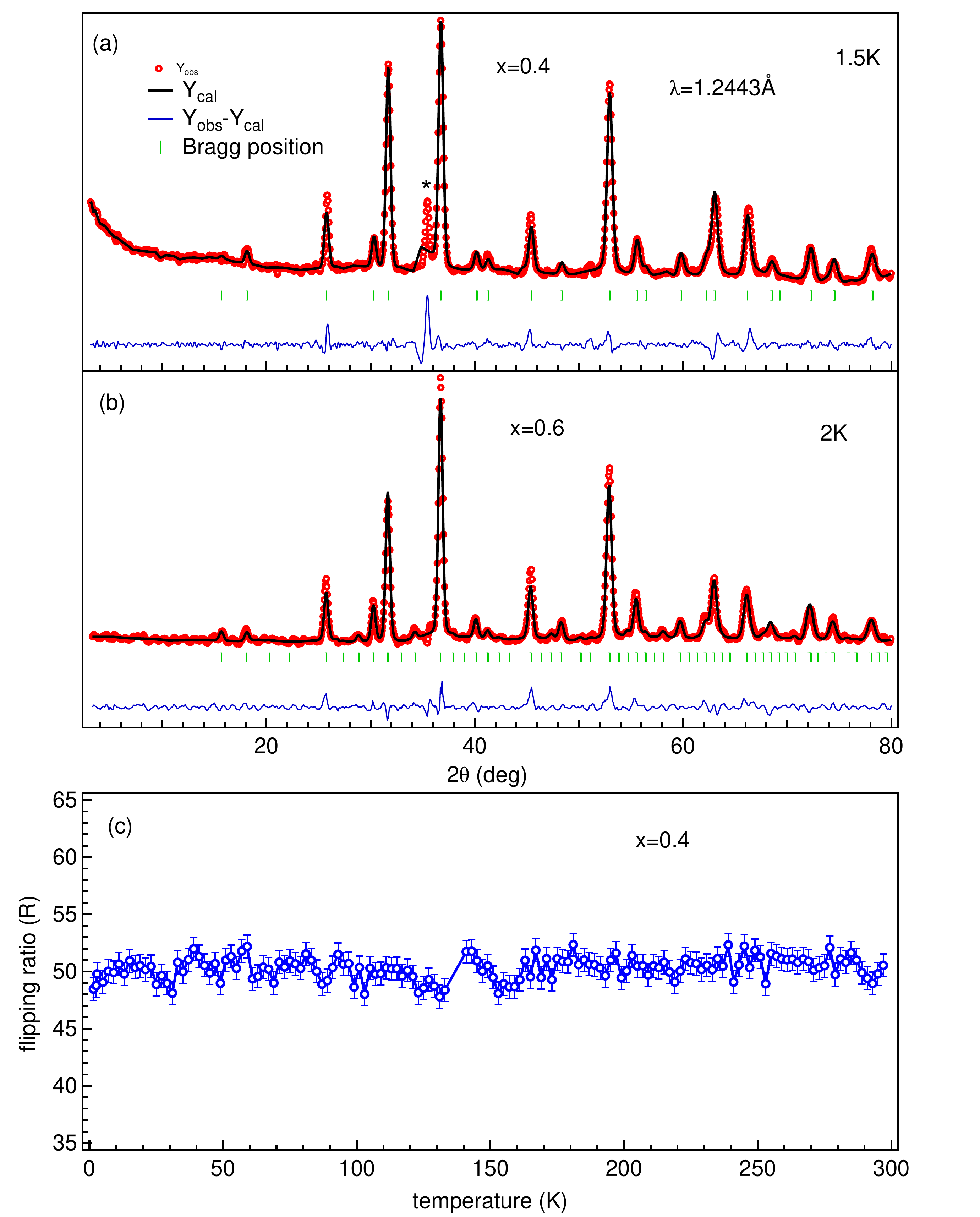}
\caption {(a) The Rietveld refinement of NPD patterns of the $x=$ 0.4 and 0.6 samples recorded at $\le$ 2~K with $\lambda =$ 1.2443 \AA. (b) The temperature dependence of the flipping ratio (R) for the $x=$ 0.4 sample.} 
\label{short_4}
\end{figure}

Finally, to further probe the possibility of the low temperature magnetic ordering, the NPD patterns are recorded at $\le$2~K up to 2$\theta=$ 3$^{\rm o}$ using $\lambda=$ 1.2443~\AA $\space$,~as shown in Fig.~\ref{short_4}(a, b). Neither the additional magnetic reflections above the background level nor any significant enhancement in the intensity of the nuclear peaks have been observed in the $x=$ 0.4 sample down to 1.5~K, suggesting the presence of short-range magnetic correlations, which is consistent with the magnetization measurements, showing the low temperature cluster-glass-like behavior for the $x=$ 0.4 sample \cite{Kumar_PRB2_20}. An additional Bragg reflection around 35$^0$ in Fig.~\ref{short_4}(a) as indicated by the asterisk symbol is originating from the pillar of cryomagnet [(111) reflection of 304L stainless steel]. The NPD measurement at 1.5~K with $\lambda=$ 1.2443~\AA~ also support the completely disordered structure of the $x=$ 0.4 sample, as evident from that measurements with $\lambda= $1.094~\AA~ presented in Figs.~\ref{all_4}(a--g)]. In order to further probe the short range ferromagnetic (FM) spin-spin correlations in the $x=$ 0.4 sample, neutron depolarization measurements were performed down to 1.5~K, as shown in Fig.~\ref{short_4}(c). The Larmor precession of the neutron spin about a net non-zero magnetic moment depolarize the incident polarized neutron beam, when it pass through the ferromagnetic spin-cluster and hence can probe the FM domains down to $\sim$100~\AA~ \cite{Manna_JPCM_14, Mitsuda_PRB_92, Halder_PRB_11, Yusuf_PRB_03}. On the other hand, canonical spin glass or AFM systems with the zero net magnetic moment can not depolarize the neutron beam due to their zero net magnetic moment \cite{Mitsuda_PRB_92, Mirebeau_PRB_90}. In the present case, no depolarization of the neutron beam is observed down to 1.5~K [see Fig.~\ref{short_4}(c)], which discard the presence of FM clusters in the sample. This is consistent with the ac magnetic susceptibility measurements, which indicate the weak spin-spin correlations in the $x =$ 0.4 sample \cite{Kumar_PRB2_20}. 

For the $x=$ 0.6 sample, the magnetization as well as specific heat measurements show the low temperature AFM interactions \cite{Kumar_PRB1_20, Kumar_PRB2_20}; however, surprisingly no additional magnetic reflections are observed in the NPD pattern measured at 2~K, as shown in the Fig.~\ref{short_4}(b) along the refinement with nuclear phase only. The presence of 32\% disorder between the Co and Nb atoms in the present case results in the additional Co--O--Co magnetic interactions, which may be the possible reason for suppression of the long range Co--O--Nb--O--Co AFM path. This suggest that the range of the AFM ordering in the $x=$ 0.6 sample is not long enough to produce the magetic reflection in the NPD spectra above the background level, as this sample lies at the borderline of the ordered-disordered configurations of Sr$_{2-x}$La$_x$CoNbO$_6$ ($x =$0--1) series \cite{Kumar_PRB1_20, Kumar_PRB2_20, BoyaPRB21}. On the other hand, the NPD pattern reported for the $x=$ 1 sample (with 4\% $B$-site disorder) clearly show the additional magnetic reflections resulting from the antiferromagnetically aligned spins with the magnetic prorogation vector $k$= ($\frac{1}{2}$ 0 $\frac{1}{2}$) \cite{Bos_PRB_04}. 

\section{\noindent ~Conclusions}

In summary, we report the temperature dependent NPD measurements on Sr$_{2-x}$La$_x$CoNbO$_6$ ($x=$0.4, 0.6) samples. Evolution of the superlattice reflection clearly evident the sudden enhancement in the B-site ordering from $x =$0.4 to 0.6 samples. Rietveld refinement  of the NPD data indicate the completely random occupancy of the Co and Nb atoms at B-site ($S=$ 0) for the $x=$ 0.4 sample, while 68\% ordered structure for the $x=$ 0.6 sample. Further, we observe a transformation from tetragonal ($I4/m$) to monoclinic ($P2_1/n$) phase from $x=$ 0.4 to 0.6. We suggest that the abrupt change in the extant B-site ordering from $x=$ 0.4 to 0.6 is responsible for this structural phase transition and the evolution of long range magnetic interactions in these compounds for the $x>$ 0.6 samples. The evolution of the unit cell volume with temperature follow the  Gr\"uneisen approximation for both the samples. We observe the anisotropic thermal expansion of the unit cell of $x= $0.6 sample, where c-axis do not follow the Gr\"uneisen approximation. A significant reduction in the degree of octahedral distortion at both Co and Nb sites has been observed in the $x=$ 0.6 sample as compared to the $x=$ 0.4 sample due to lower concentration of JT active Co$^{3+}$ ions in the former. Moreover, an increase in the local distortion around $A$-site atoms (La/Sr) is evident with increase in the La substitution from $x=$ 0.4 to $x=$ 0.6 sample. The absence of additional magnetic reflections down to $\le$ 2~K indicates the short range magnetic interactions in these samples.\\
\\

{\bf \large ACKNOWLEDGMENTS}\\

AK thanks the UGC for fellowship through IIT Delhi. We acknowledge the financial support from SERB-DST through core research grant (file no.: CRG/2020/003436). The powder neutron diffraction measurements were performed under CRS project no. UDCSR/MUM/AO/CRS-M-270/2017/567. \\
\\


\begin{thebibliography}{99}

\bibitem{Perez_PRL_98} P\'erez, J.; Garc\'ia, J.; Blasco, J.; Stankiewicz, J. Spin-glass behavior and giant magnetoresistance in the (RE)Ni$_{0.3}$Co$_{0.7}$O$_3$ (RE=La, Nd, Sm) system. Phys. Rev. Lett. {\bf 1998}, 80, 2401-2404.

\bibitem{Kundu_PRB_05} Kundu, A. K.; Nordblad, P.; Rao, C. N. R. Nonequilibrium magnetic properties of single-crystalline La$_{0.7}$Ca$_{0.3}$CoO$_3$. Phys. Rev. B {\bf 2005}, 72, No. 144423.

\bibitem{Giovannetti_PRL_09} Giovannetti, G.; Kumar, S.; Khomskii, D.; Picozzi, S.; Brink, J. V. D. Multiferroicity in rare-earth nickelates RNiO$_3$. Phys. Rev. Lett. {\bf 2009}, 103, No. 156401.

\bibitem{Sheng_PRB_09} Sheng, Z. G.; Gao, J.; Sun, Y. P. Current-induced resistive switching effect in oxygen-deficient La$_{0.8}$Ca$_{0.2}$MnO$_{3-\delta}$ films. Phys. Rev. B {\bf 2009}, 79, No. 014433.

\bibitem{Tao_NM_03} Tao, S.; Irvine, J. T. S. A redox-stable efficient anode for solid-oxide fuel cells. Nat. Mater. {\bf 2003}, 2, 320-323.

\bibitem{Chakrabartty_NP_18} Chakrabartty, J.; Harnagea, C.; Celikin, M.; Rosei, F.; Nechache, R. Improved photovoltaic performance from inorganic perovskite oxide thin films with mixed crystal phases. Nat. Photonics. {\bf 2018}, 12, 271-276.

\bibitem{Das_PRB_17} Das, M.; Roy, S.; Mandal, P. Giant reversible magnetocaloric effect in a multiferroic GdFeO$_3$ single crystal. Phys. Rev. B {\bf 2017}, 96, No. 174405. 

\bibitem{Anderson_SSC_93} Anderson, M. T.; Greenwood, K. B.; Taylor, G. A.; Poeppelmeier, K. R. B-cation arrangements in double perovskites. Prog. in solid st. Chem. {\bf 1993}, 22, 197-232.

 \bibitem{King_JMC_10} King, G.; Woodward, P. M. Cation ordering in perovskites. J. Mater. Chem. {\bf 2010}, 20, 5785-5796.

  \bibitem{Vasala_SSC_15} Vasala. S.; Karppinen, M. A$_2$B'B''O$_6$  perovskites: A review.  
Prog. in solid st. Chem. {\bf 2015}, 43, 1-36.

 \bibitem{Galasso_JPC_62} Galasso, F.; W. Darby, W. Ordering of the octahedrally coordinated cation position in the perovskite structure. J. Phys. Chem. {\bf 1962}, 66, 131-132.

\bibitem{Niebieskikwiat_PRB_04} Niebieskikwiat, D.; Prado, F.; Caneiro, A.; S\'anchez, R. D. Antisite defects versus grain boundary competition in the tunneling magnetoresistance of Sr$_2$FeMoO$_6$ double perovskite. Phys. Rev. B {\bf 2004}, 70, No. 132412.

\bibitem{Sanchez_PRB_02} S\'anchez, D.; Alonso, J. A.; Garc\'ia-Hern\'andez, M.; Mart\'inez-Lope, M. J.; Mart\'inez, J. L.; Mellerg\aa rd, A. Origin of neutron magnetic scattering in antisite-disordered Sr$_2$FeMoO$_6$ double perovskites. Phys. Rev. B {\bf 2002}, 65, No. 104426.

\bibitem{Jung_PRB_07} Jung, A.; Bonn, I.; Ksenofontov, V.; Panth\"ofer, M.; Reiman, S.; Felser, C.; Tremel, W. Effect of cation disorder on the magnetic properties of Sr$_2$Fe$_{1-x}$Ga$_x$ReO$_6$ (0\textless$x$\textless0.7) double perovskites. Phys. Rev. B {\bf 2007}, 75, No. 184409. 

\bibitem{Erten_PRL_11} Erten, O.; Meetei, O. N.; Mukherjee, A.; Randeria, M.; Trivedi, N.; Woodward, P. Theory of half-metallic ferrimagnetism in double  perovskites. Phys. Rev. Lett. {\bf 2011}, 107, No. 257201.

\bibitem{Frontera_PRB_04} Frontera,  C.; Fontcuberta, J. Configurational disorder and magnetism in double perovskites: A Monte Carlo simulation study. Phys. Rev. B {\bf 2004}, 69, No. 014406.

\bibitem{Saines_JSSC_07} Saines, P. J.; Spencer, J. R.; Kennedya, B. J.; Avdeevb, M. Structures and crystal chemistry of the double perovskites Ba$_2$LnB$^\prime$O$_6$ (Ln= lanthanide B$^\prime$= Nb$^{5+}$ and Ta$^{5+}$): Part I. Investigation of Ba$_2$LnTaO$_6$ using synchrotron X-ray and neutron powder diffraction. J. Solid State Chem. {\bf 2007}, 180, 2991-3000.

\bibitem{Kumar_PRB1_20} Kumar, A.; Dhaka, R. S. Unraveling magnetic interactions and the spin state in insulating Sr$_{2-x}$La$_x$CoNbO$_6$. Phys. Rev. B {\bf 2020}, 101, No. 094434.

 \bibitem{Bos_PRB_04} Bos, J.-W. G.; Attfield, J. P. Magnetic frustration in La(A)CoNbO$_6$ (A=Ca, Sr, and Ba) double perovskites. Phys. Rev. B {\bf 2004}, 70, No. 174434.

\bibitem{Lloret_ICA_08} Lloret, F.; Julve, M.; Cano, J.; Ruiz-Garc\'ia, R.; Pardo, E. Magnetic properties of six-coordinated high-spin Co(II) complexes: Theoretical background and its application. Inorg. Chem. Acta {\bf 2008}, 361, 3432-3445. 

\bibitem{Shukla_PRB_18} Shukla, R.; and Dhaka, R. S. Anomalous magnetic and spin glass behavior in Nb substituted LaCo$_{1-x}$Nb$_x$O$_3$. Phys. Rev. B {\bf 2018}, 97, No. 024430.

\bibitem{Shukla_JPCC_19} Shukla, R.; Jain, A.; Miryala, M.; Murakami, M.; Ueno, K.; Yusuf, S. M.; Dhaka, R. S. Spin dynamics and unconventional magnetism in insulating La$_{(1-2x)}$Sr$_{2x}$Co$_{(1-x)}$Nb$_x$O$_3$. J. Phys. Chem. C {\bf 2019}, 123, 22457-22469.

\bibitem{Kumar_PRB2_20} Kumar, A.; Schwarz, B.; Ehrenberg, H.; Dhaka, R. S. Evidence of discrete energy states and cluster-glass behavior in Sr$_{2-x}$La$_x$CoNbO$_6$. Phys. Rev. B {\bf 2020}, 102, No. 184414.

\bibitem{Sow_PRB_12} Sow, C.; Samal, D.; Kumar, P. S. A.; Bera, A. K.; Yusuf, S. M. Structural-modulation-driven low-temperature glassy behavior in SrRuO$_3$. Phys. Rev. B {\bf 2012}, 85, No. 224426.

\bibitem{Carvajal_PB_93} Rodr\'iguez-Carvajal, J. Recent advances in magnetic structure determination by neutron powder diffraction. Physica B {\bf 1993}, 192, 55-69.

\bibitem{Yusuf_Pramana_96} Yusuf, S. M.; Rao L. M. Magnetic studies in mesoscopic length scale using polarized neutron spectrometer at Dhruva reactor, Trombay. Pramana J. Phys. {\bf 1996}, 47, 171-182.

\bibitem{Zhu_PRB_20} Zhu, Y.; Wu, S.; Tu, B.; Jin, S.; Huq, A.; Persson, J.; Gao, H.; Ouyang, D.; He, Z.; Yao, D.-X.; Tang, Z.; Li, H.-F. High-temperature magnetism and crystallography of a YCrO$_3$ single crystal. Phys. Rev. B {\bf 2020}, 101, No. 014114.

\bibitem{Wallace_book_98} Wallace, D. C. Thermodynamics of Crystals (Dover, New York, {\bf 1998}).

\bibitem{Shannon_AC_76} Shannon, R. D. Revised effective ionic radii and systematic studies of interatomic distances in halides and chalcogenides. Acta. Cryst. {\bf 1976}, A32, 751-767. 

\bibitem{Chen_JACS_14} Chen, J. -M.; Chin, Y.-Y.; Valldor, M.; Hu, Z.; Lee, J.-M.; Haw, S.-C.; Hiraoka, N.; Ishii, H.; Pao, C.-W.; Tsuei, K.-D.; Lee, J.-F.; Lin, H.-J.; Jang, L.-Y.; Tanaka, A.; Chen, C.-T.; Tjeng, L. H. A complete high-to-low spin state transition of trivalent cobalt ion in octahedral symmetry in SrCo$_{0.5}$Ru$_{0.5}$O$_{3-\delta}$. J. Am. Chem. Soc. {\bf 2014},136, 1514-1519.

\bibitem{Shukla_JPCC_21} Shukla, R.; Kumar, A.; Kumar, R.; Jha, S. N.; Dhaka, R. S. X-ray absorption spectroscopy study of La$_{1-y}$Sr$_y$Co$_{1-x}$Nb$_x$O$_3$. J. Phys. Chem. C {\bf 2021}, 125,  10130-10139.

\bibitem{Pradheesh_EPJB_12} Pradheesh, R.; Nair, H. S.; Sankaranarayanan, V.; Sethupathi, K. Large magnetoresistance and Jahn-Teller effect in Sr$_2$FeCoO$_6$. Eur. Phys. J. B {\bf 2012}, 85, No. 260.

\bibitem{BaiIC19} Bai, Y.; Han, L.; Meng, J.;  Baran, V.;  Hao, J.; Han, L. Connection between unusual lattice thermal expansion and cooperative Jahn-Teller effect in double perovskites LaPbMSbO$_6$ (M= Mn, Co, Ni). Inorg. Chem. {\bf 2019}, 58, 2888-2898.

\bibitem{Kumar_XAS_22} Kumar, A.; Shukla, R.; Kumar, R.; Choudhary, R. J.; Jha, S. N.; Dhaka, R. S. Probing the electronic and local structure of Sr$_{2-x}$La$_x$CoNbO$_6$ using near-edge and extended x-ray absorption fine structures, unpublished.

\bibitem{Manna_JPCM_14} Manna, K.; Samal, D.; Bera, A. K.; Elizabeth, S.; Yusuf, S. M.; Kumar, P. S. A. Correspondence between neutron depolarization and higher order magnetic susceptibility to investigate ferromagnetic clusters in phase separated systems. J. Phys.: Condens. Matter {\bf 2014}, 26, No. 016002. 

\bibitem{Mitsuda_PRB_92} Mitsuda, S.; Yoshizawa, H.; Endo, Y. Neutron-depolarization studies on re-entrant spin glass. Phys. Rev. B {\bf 1992}, 45, No. 9788.

\bibitem{Halder_PRB_11}Halder, M.; Yusuf, S. M.; Kumar, A.; Nigam, A. K.; Keller, L. Crossover from antiferromagnetic to ferromagnetic ordering in the semi-Heusler alloys Cu$_{1-x}$Ni$_x$MnSb with increasing Ni concentration. Phys. Rev. B {\bf 2011}, 84, No. 094435.

\bibitem{Yusuf_PRB_03} Yusuf, S. M.; Chakraborty, K. R.; Paranjpe, Ganguly, R.; Mishra, P. K.;  Yakhmi, J. V.; Sahni, V. C. Magnetic and electrical properties of (La$_{1-x}$Dy$_x$)$_{0.7}$Ca$_{0.3}$MnO$_3$ perovskites. Phys. Rev. B {\bf 2003}, 68, No. 104421.

\bibitem{Mirebeau_PRB_90} Mirebeau, I.; Yusuf, Itoh, Mitsuda, S.; Watanabe, T.; Endoh, Y.; Hennion, M.;  Papoular, R. Neutron depolarization in a reentrant spin-glass system: Amorphous Fe-Mn. Phys. Rev. B {\bf 1990}, 41, No. 11405.

\bibitem{BoyaPRB21} Boya,  K.; Nam, K.; Manna, A. K.; Kang, J.; Lyi, C.; Jain, A.; Yusuf, S. M.; Khuntia, P.; Sana, B.; Kumar, V.; Mahajan, A. V.; Patil, D. R.; Kim, K. H.; Panda, S. K.; Koteswararao, B. Magnetic properties of the $S=$ 5/2 anisotropic triangular chain compound Bi$_3$FeMo$_2$O$_{12}$. Phys. Rev. B {\bf 2021}, 104, No. 184402.

\end{thebibliography}
\end{document}